\pdfoutput=1
\documentclass[10pt, conference, compsocconf, letter]{IEEEtran}

\usepackage{ntheorem}

\theoremstyle{plain}
\theorembodyfont{\normalfont}
\newtheorem{example}{Example}
\newtheorem{theorem}{Theorem}
\newtheorem{lemma}{Lemma}
\newtheorem{definition}{Definition}

\usepackage{proof,bbm,mdwtab}
\usepackage{calc}

\usepackage{cite}

\ifCLASSINFOpdf
\else
\fi

\usepackage[cmex10]{amsmath}
\usepackage{amssymb}

\usepackage{listings,color,textcomp}

\usepackage{tikz}
\usetikzlibrary{arrows,positioning,fit,shapes,calc,shapes.multipart}

\tikzset{depSort/.style={draw=black!50,line width=1pt,rounded corners=2pt},
         depCtor/.style={},
         linkStyle/.style={line width = 0.6}}

\newcommand{\ExternalLink}[1]{%
  \href{#1}{%
    \tikz[x=1.2ex, y=1.2ex, baseline=-0.05ex]{%
        \begin{scope}[x=1ex, y=1ex]
            \clip (-0.1,-0.1) 
                --++ (-0, 1.2) 
                --++ (0.6, 0) 
                --++ (0, -0.6) 
                --++ (0.6, 0) 
                --++ (0, -1);
            \path[draw, linkStyle, 
                rounded corners=0.5] 
                (0,0) rectangle (1,1);
        \end{scope}
        \path[draw, linkStyle] (0.5, 0.5) 
            -- (1, 1);
        \path[draw, linkStyle] (0.6, 1) 
            -- (1, 1) -- (1, 0.6);
        }
    }}

\newcommand{\exLink}[1]{\makebox[0mm][l]{\hspace*{0.96\linewidth}{\exNum{#1}}}}

\newcommand{\exNum}[1]{\ExternalLink{http://logicrunch.it.uu.se:4096/~wv/princess/?ex=ADTs\%2Fex\%3A#1}}

\usepackage{url,hyperref}

\hypersetup{
    colorlinks=false,
    pdfborder={0 0 0},
}

\newcommand{\id}[1]{\ensuremath{\texttt{#1}}}
\newcommand{\Z}{\ensuremath{\mathbbm{Z}}}
\newcommand{\N}{\ensuremath{\mathbbm{N}}}
\newcommand{\T}{\ensuremath{\mathbbm{T}}}
\newcommand{\BBS}{\ensuremath{\mathbbm{S}}}

\newcommand{\val}{\ensuremath{\operatorname{val}}}

\newif\ifcomments
\commentsfalse

\newif\ifoutline
\outlinefalse

\newif\iflong
\longtrue

\newcommand{\Comment}[1]{\ifcomments\marginpar{\small\color{blue} #1}\fi}
\newcommand{\contents}[1]{\ifoutline{\color{blue}
    \begin{itemize}
    #1
    \end{itemize}
  }\fi}

\begin{document}
\title{Deciding and Interpolating Algebraic Data Types by Reduction
  \iflong(Technical Report)\fi}


\author{\IEEEauthorblockN{Hossein Hojjat}
\IEEEauthorblockA{Department of Computer Science\\
 Rochester Institute of Technology,
Rochester, NY, United States
}
\and
\IEEEauthorblockN{Philipp R\"ummer}
\IEEEauthorblockA{Department of Information Technology\\
  Uppsala University,
  Uppsala, Sweden
}
}


%


\maketitle

\begin{abstract}
  Recursive algebraic data types (term algebras, ADTs) are one of the
  most well-studied theories in logic, and find application in
  contexts including functional programming, modelling languages,
  proof assistants, and verification. At this point, several
  state-of-the-art theorem provers and SMT solvers include tailor-made
  decision procedures for ADTs, and version~2.6 of the SMT-LIB
  standard includes support for ADTs. We study an extremely simple
  approach to decide satisfiability of ADT constraints, the reduction
  of ADT constraints to equisatisfiable constraints over uninterpreted
  functions (EUF) and linear integer arithmetic (LIA). We show that
  the reduction approach gives rise to both decision and Craig
  interpolation procedures in (extensions of) ADTs.
\end{abstract}


%
\IEEEpeerreviewmaketitle

\section{Introduction}

Recursive algebraic data types (ADTs) with absolutely free
constructors are increasingly supported by SMT solvers, and find
application in a variety of areas, including functional programming,
modelling languages, proof assistants, and verification. In solvers,
ADTs are usually implemented as native theory
solvers~\cite{Oppen:1980:RRD:322203.322204,DBLP:journals/jsat/BarrettST07,Suter:2010:DPA:1707801.1706325,DBLP:journals/jar/ReynoldsB17}
that apply congruence closure (upward closure), syntactic unification
(downward closure), cycle detection (occurs-check), and in additional
handle selectors and testers in order to decide satisfiability of
quantifier-free ADT formulas.

In this paper, we study a simple alternative approach to ADT
reasoning, based on the \emph{reduction} of ADT formulas to
equisatisfiable formulas over uninterpreted functions and linear
integer arithmetic (EUF+LIA).
Our approach is partly inspired, besides by eager SMT in general, by
the reduction approach from \cite{kapurInt}, in which quantifier-free
formulas are mapped to simpler theories for the purpose of checking
satisfiability and computing interpolants. For instance, as shown in
\cite{kapurInt}, the theory of sets with finite cardinality
constraints can be reduced to the theory of equality with
uninterpreted functions (EUF). Like in \cite{kapurInt}, the target
theories of our ADT reduction are EUF and linear arithmetic. Unlike
\cite{kapurInt}, we are able to completely avoid universal quantifiers
in the process of reduction, but the reduction depends on the
introduction of further uninterpreted functions (which create some
additional work in interpolation, see
Section~\ref{sec:interpolation}).

The main idea of reduction is to augment an ADT formula with
additional literals that ensure that constructors, selectors, and
testers are interpreted consistently, and that constructors are free.
EUF takes care of upward and downward closure, while cycle detection
and constructor testers are handled by LIA constraints. The reduction can
be implemented with little effort, and is widely applicable since EUF
and LIA are supported by virtually all SMT solvers, and increasingly
also by other theorem provers. Reduction to EUF+LIA has a few further
advantages, in particular it is possible to reuse existing, highly
optimised EUF+LIA simplifiers in solvers, and to compute interpolants
using EUF+LIA interpolation procedures.

The contributions of the paper are
(i)~definition and correctness proof of the reduction from ADTs to
EUF+LIA;
(ii)~discussion of Craig interpolation for ADTs;
(iii)~extension to ADTs with size constraints, and an effective
characterisation of the ADTs for which the resulting procedure is
complete.
The procedures discussed in the paper have been implemented in the
\textsc{Princess} theorem
prover~\cite{princess08}.\footnote{\texttt{\url{http://www.philipp.ruemmer.org/princess.shtml}}}


\subsection{Related Work}

\paragraph{ADT Solving}
While ADTs have only recently been standardised in the SMT-LIB, some
solvers (including STeP~\cite{DBLP:conf/tapsoft/MannaBBCCADKLSU95},
CVC3~\cite{BT07}, CVC4~\cite{DBLP:conf/cav/BarrettCDHJKRT11}, and
Z3~\cite{DBLP:conf/tacas/MouraB08}) have for a while supported ADTs
through native decision procedures extending the congruence closure
algorithm~\cite{Oppen:1980:RRD:322203.322204,DBLP:journals/jsat/BarrettST07,DBLP:journals/jar/ReynoldsB17}.
Native solvers offer excellent performance, but also require
significant implementation effort. The listed solvers do not support
Craig interpolation or formulas with size constraints.

Satisfiability of ADT formulas can also be checked by introducing
explicit axioms about the constructors and selectors. Since ADTs form
a local theory~\cite{DBLP:conf/cade/Sofronie-Stokkermans09}, the set
of required instances of the axioms can effectively be computed, and a
decision procedure for ADT satisfiability is obtained.

Our reduction-based approach sits in between native solvers and
methods based on explicit axioms. Like with explicit axioms, our
method leaves most of the heavy work to other theory solvers (EUF and
LIA), and is therefore easy to implement. The reduction approach is
structure-preserving, however, which makes us believe that it can
utilise existing contextual simplifiers (pre-processors or
in-processors) more effectively than approaches based on axioms; it
also directly gives rise to an interpolation procedure.

\paragraph{ADT Interpolation}
It has been observed in \cite{kapurInt} that the theory of ADTs has
the interpolation property; this result directly follows from
admissibility of quantifier elimination in
ADTs~\cite{Oppen:1980:RRD:322203.322204}. To the best of our
knowledge, our ADT solver implemented in \textsc{Princess} is the
\emph{first proof-based interpolation procedure for ADTs.}

\paragraph{ADTs with Size Constraints}
Our approach for handling ADT formulas with size constraints is
inspired by the more general unfolding-based decision procedure for
ADTs with abstractions (i.e., catamorphisms) in
\cite{Suter:2010:DPA:1707801.1706325}. The algorithm in
\cite{Suter:2010:DPA:1707801.1706325} is complete for
\emph{sufficiently surjective} abstraction functions, which includes
the size function on binary trees, but \emph{not} the size function on
ADTs in general. We augment the setting from
\cite{Suter:2010:DPA:1707801.1706325} by giving a necessary and
sufficient criterion for sufficient surjectivity of the size function,
and thus for completeness of the overall procedure.

ADTs with size constraints can also be represented in the local theory
framework~\cite{DBLP:conf/cade/Sofronie-Stokkermans09}, again by
introducing the necessary instances of explicit axioms.

A further decision procedure for ADTs with size constraints, based on
the concept of length constraint completion, has been described in
\cite{DBLP:journals/iandc/ZhangSM06}. Our method uses the simple
approach of \emph{unfolding} in order to add size constraints to the
overall reduction-based procedure; it is at this point unclear whether
length constraint completion could be combined with the reduction
approach as well.


\section{Preliminaries}
\label{sec:prelims}



We formulate our approach in the setting of multi-sorted first-order
logic.  The signature~$\Sigma$ of an ADT is defined by a
sequence~$\sigma^d_1, \ldots, \sigma^d_k$ of sorts and a
sequence~$f_1, \ldots, f_m$ of constructors.  The type~$\alpha(f_i)$
of an $n$-ary constructor is an $(n+1)$-tuple~$\langle\sigma_0,
\ldots, \sigma_n\rangle \in \{\sigma^d_1, \ldots, \sigma^d_k\}^{n+1}$,
normally written in the form~$f_i : \sigma_1 \times \cdots \times
\sigma_n \to \sigma_0$. Zero-ary constructors are also called
constants. By slight abuse of notation, we also write $f_i :
\sigma^d_j$ if the result type of $f_i$ is $\sigma^d_j$, i.e., if $f_i
: \sigma_1 \times \cdots \times \sigma_n \to \sigma^d_j$ for some
$\sigma_1, \ldots, \sigma_n$.

In addition to constructors, formulas over ADTs can be formulated in
terms of \emph{variables}~$x \in \cal X$ (with some type $\alpha(x)
\in \{\sigma^d_1, \ldots, \sigma^d_k\}$); \emph{selectors}~$f_i^j$,
which extract the $j^{\text{th}}$ argument of an $f_i$-term; and
\emph{testers}~$\mathit{is}_{f_i}$, which determine whether a term is
an $f_i$-term. The syntactic categories of terms~$t$ and
formulas~$\phi$ are defined by the following rules:
\begin{align*}
  t ~~::= & ~~ x && \text{Variables}
  \\
  | ~ & ~~ f_i(\bar t) && \text{Constructors}
  \\
  | ~ & ~~ f_i^{j}(t) && \text{Selectors}
  \\
  \phi ~~::= & ~~ \mathit{is}_{f_i}(t) && \text{Testers}
  \\
  |~ & ~~ t \approx t && \text{Equality}
  \\
  |~ & ~~ \phi \wedge \phi ~|~ \phi \vee \phi ~|~ \neg \phi ~|~ \ldots
                           &&
                              \text{Boolean operators}
\end{align*}

Well-typed terms and formulas are defined as expected, assuming that
selectors have type $f_i^j : \sigma_0 \to \sigma_j$ whenever $f_i : \sigma_1
\times \cdots \times \sigma_n \to \sigma_0$, and
testers~$\mathit{is}_{f_i}$ expect an argument of type~$\sigma^d_j$ if
$f_i : \sigma^d_j$. In the whole paper, we assume that considered
expressions are well-typed.

\begin{example}[Lists]
  \label{ex:lists}
  We show examples in the concrete syntax used in our implementation.
\begin{lstlisting}[escapechar=\%]
\sorts {
  Colour { red; green; blue; };
  CList { nil; cons(Colour head, CList tail); };
}
\end{lstlisting}

Given variables~$x$ of sort~\id{CList} and $y$ of sort~\id{Colour}, a
formula over this data type is:\iflong\footnote{In all examples, the link
  \exNum{1} will take you to the web interface of our SMT solver
  \textsc{Princess} and directly load the given constraint.}\fi
\begin{lstlisting}[escapechar=\%]
%\exLink{1}%  x.is_cons & y != blue &
  (x.head = red | x = cons(y, nil))
\end{lstlisting}
corresponding to the abstract syntax formula
\begin{align*}
  &\mathit{is}_{\id{cons}}(x) \wedge \neg(y \approx \id{blue})  \wedge~
  \\
  &\big(\id{cons}^1(x) \approx \id{red} \vee
  x \approx \id{cons}(y, \id{nil}) \big)
\end{align*}
Assigning $\{x \mapsto \id{cons}(\id{red}, \id{nil}), y\mapsto \id{green} \}$
satisfies the formula.
\hfill\IEEEQED
\end{example}

A constructor term is a ground term~$t$ that only consists of
constructors (i.e., does not contain selectors or variables). We
denote the set of all constructor terms (for some fixed ADT signature)
by $\T$, and the set of all constructor terms of
type~$\sigma^d_j$ by $\T_{\sigma^d_j}$. An ADT is well-defined if
$\T_{\sigma^d_j}$ is non-empty for all sorts~$\sigma^d_1, \ldots,
\sigma^d_k$, and we will henceforth only consider well-defined ADTs.

Semantics is defined in terms of structures~$(\T, I)$ over the
universe~$\T$ of constructor terms, i.e., constructors are absolutely
free. Selectors~$f^j : \sigma_0 \to \sigma_j$ in particular are mapped
to total set-theoretic functions
$I(f^j) : \T_{\sigma_0} \to \T_{\sigma_j}$ satisfying
$I(f^j)(f(t_1, \ldots, t_n)) = t_j$.



\section{A Verification Example}

As a high-level example, we outline how a simple program operating on
the ADT from Example~\ref{ex:lists} can be verified using our
procedures.
We represent the program in the form of (constrained) Horn clauses,
following the approach taken in several recent verification
systems~\cite{andrey-pldi, Rummer13}. The result resembles a classical
logic program implementing the concatenation of two lists;
\lstinline!C(x, y, r)! expresses that \lstinline!r! is the result of
concatenating lists~\lstinline!x, y!:
\begin{lstlisting}[escapechar=\%]
\forall CList y;                         // (C1)
  C(nil, y, y)
\forall CList x, y, r; \forall Colour c; // (C2)
  (C(x, y, r) -> C(cons(c, x), y, cons(c, r)))
\end{lstlisting}

As a first property of the program, we can observe that the head of a
non-empty result list~\lstinline!r! has to be the head of one of the
arguments~\lstinline!x, y!:
\begin{lstlisting}[escapechar=\%]
\forall CList x, y, r; (                 // (P1)
  r != nil & C(x, y, r) ->
  (r.head = x.head | r.head = y.head))
\end{lstlisting}

To verify this property, it is enough to find a \emph{model} of the
(constrained) Horn clauses \lstinline!(C1), (C2), (P1)!, i.e., an
interpretation of the predicate~\lstinline!C! that satisfies all three
formulas. The predicate~\lstinline!C! can then be considered as a
post-condition (or inductive invariant) that is sufficient to show
property~\lstinline!(P1)!. One solution of
\lstinline!(C1), (C2), (P1)!
is to interpret \lstinline!C(x, y, r)! as
\begin{lstlisting}[escapechar=\%]
%\exLink{P1}%C(CList x, CList y, CList r) {
  r = y | r.head = x.head
}
\end{lstlisting}
which can indeed be observed to satisfy all three clauses. The
decision procedure for ADTs defined in the next section can \iflong easily\fi
check correctness of this model mechanically, after inlining the 
definition of \lstinline!C!, and skolemising away quantifiers.

To find models of clauses like \lstinline!(C1), (C2), (P1)!
automatically, the principle of \emph{Craig interpolation} can be
applied to derivation trees of the clauses, an approach that has been
implemented in several model checkers~\cite{andrey-pldi, Rummer13}. To
support ADTs, which are currently beyond the scope of most model
checkers, in Section~\ref{sec:interpolation} we explain how our
decision procedure can be extended to Craig interpolation.

\medskip
Consider now additional clauses computing the list length:
\begin{lstlisting}[escapechar=\%]
L(nil, 0)                                // (C3)
\forall CList x;                         // (C4)
  \forall Colour c; \forall int n;
    (L(x, n) -> L(cons(c, x), n+1))
\end{lstlisting}

We can combine the two programs to state a second property relating
concatenation and list length. Concatenating two lists yields a list
whose length is the sum of the individual list lengths:
\begin{lstlisting}[escapechar=\%]
\forall CList x, y, r;                   // (P2)
  \forall int nx, ny, nr; (
    C(x, y, r) & L(x, nx) & L(y, ny) & L(r, nr)
    -> nr = nx + ny)
\end{lstlisting}

To verify this property, as before by showing the existence of a model
of \lstinline!(C1), (C2), (C3), (C4), (P2)!, we need a slightly
extended logic providing also an operator for the \emph{size} of ADT
terms (Section~\ref{sect::size}). ADT constraints without size
operator are not sufficiently expressive to formulate any model.  The
size of a term~$t \in \T$ is the number of constructor occurrences in
$t$. A model of \lstinline!(C1), (C2), (C3), (C4), (P2)!,
interpreting both the predicate
\lstinline!C! and \lstinline!L!, is then
\begin{lstlisting}[escapechar=\%]
%\exLink{P2}%C(CList x, CList y, CList r) {
  \size(x) + \size(y) = \size(r) + 1
};
L(CList x, int n) {
  \size(x) = 2*n + 1
};
\end{lstlisting}
Note that the \lstinline!\size! operator also counts the
\lstinline!nil! symbol, as well as the colour constructors
\lstinline!red, green, blue!, leading to the stated relationship
between the size and the length of a list. The correctness of the
model can be checked using the procedure we define in
Section~\ref{sect::size}.


\begin{table*}[t]
  \fboxsep2mm
  \fbox{
    \begin{minipage}{\linewidth-2\fboxsep}
      Reduction rules for
      constructors~$f : \sigma_1 \times \cdots \times \sigma_n \to
      \sigma_0$,
      selectors~$f^j : \sigma_0 \to \sigma_j$, testers~$\mathit{is}_f$,
      and equations between variables:
      \begin{align}
        \label{rule:1}
        \hspace*{10ex}
        f(x_1, \ldots, x_n) \approx x_0
        \quad\Longrightarrow\quad &
        \mathit{CtorSpec}_f(\tilde x_0, \ldots, \tilde x_n)
        \\
        \label{rule:2}
        f^j(x) \approx y
        \quad\Longrightarrow\quad &
        \tilde f^j(\tilde x) \approx \tilde y ~\wedge~
        \bigvee_{\substack{g \in \{f_1, \ldots, f_m\} \\ g : \sigma_0}}
        \mathit{ExCtorSpec}_g(\tilde x)
        \\
        \label{rule:3}
        \mathit{is}_f(x)
        \quad\Longrightarrow\quad &
        \mathit{ExCtorSpec}_f(\tilde x)
        \\
        \label{rule:4}
        \neg\mathit{is}_f(x)
        \quad\Longrightarrow\quad &
        \bigvee_{\substack{g \in \{f_1, \ldots, f_m\} \\ g : \sigma_0 \text{~and~} g \not= f}}
        \mathit{ExCtorSpec}_g(\tilde x)
        \\
        \label{rule:5}
        x \approx y
        \quad\Longrightarrow\quad &
        \tilde x \approx \tilde y
        \\
        \label{rule:6}
        \neg(x \approx y)
        \quad\Longrightarrow\quad &
        \neg(\tilde x \approx \tilde y)
      \end{align}

      The following abbreviations are used, for each
      constructor $f : \sigma_1 \times \cdots \times \sigma_n \to \sigma_0$
      and each sort~$\sigma \in \{\sigma^d_1, \ldots, \sigma^d_k\}$:
      \begin{align*}
      \mathit{CtorSpec}_f(x_0, \ldots, x_n)
    \quad=\quad &
    \left(
    \begin{array}{@{}>{\displaystyle}l@{}}
      \tilde f(x_1, \ldots, x_n) \approx x_0 \wedge
      \mathit{ctorId}_{\sigma_0}(x_0) \approx \mathit{Id}_f ~\wedge
      \\
      \bigwedge_{j = 1}^n \big(
      \tilde f^j(x_0) \approx x_j \wedge
      \mathit{depth}_{\sigma_0}(x_0) > \mathit{depth}_{\sigma_j}(x_j)
      \big)
    \end{array}
    \right)
    \\[1ex]
    \mathit{ExCtorSpec}_f(x)
    \quad=\quad &
    \exists x_1, \ldots, x_n.~\Big(
    \bigwedge_{j = 1}^n
    \mathit{In}_{\sigma_j}(x_j) \wedge
    \mathit{CtorSpec}_f(x, x_1, \ldots, x_n)
    \Big)
    \\[1ex]
    \mathit{In}_\sigma(x)
    \quad=\quad &
    \begin{cases}
      0 \leq x < |\T_{\sigma}| &
      \text{if~} |\T_{\sigma}| < \infty
      \\
      \mathit{true} & \text{otherwise}
    \end{cases}
  \end{align*}
    \end{minipage}}

  \caption{Rules for reduction of ADTs to EUF+LIA}
  \label{tab:reduction}
\end{table*}


\section{Checking ADT Satisfiability by Reduction}


We now define our reduction from ADTs to EUF+LIA. Suppose $\phi$ is an
ADT formula as defined in Section~\ref{sec:prelims}. For sake of
presentation, we assume that $\phi$ has been brought into a
\emph{flat} form upfront. A formula~$\phi$ is flat if function symbols
(in our case, constructors and selectors) only occur in equations of
the form~$g(x_1, \ldots, x_n) \approx x_0$ (where $x_0, \ldots, x_n$
are variables, though not necessarily pairwise distinct), and only in
positive positions. Flatness can be established at the cost of
introducing a linear number of additional variables.

\begin{example}
  \label{ex:flattened}
  The formula in Example~\ref{ex:lists} can be flattened by
  introducing variables~$\id{t1}, \id{t2} :
  \id{Colour}$, and $\id{t3} : \id{CList}$:
\begin{lstlisting}[escapechar=\%]
%\exLink{2}%  x.is_cons & blue = t1 & y != t1 &
  ((red = t2 & x.head = t2) |
   (nil = t3 & cons(y, t3) = x))
\end{lstlisting}
\end{example}

\paragraph{Notation}
We need some further notation before we can formally define the
reduction. As before, we assume that $k$ sorts $\sigma^d_1, \ldots,
\sigma^d_k$ and $m$ constructors $f_1, \ldots, f_m$ have been fixed.
For each sort~$\sigma \in \{\sigma^d_1, \ldots, \sigma^d_k\}$, we define
$\mathit{\#Ctor}_\sigma$ to be the number of constructors of $\sigma$:
\begin{equation*}
  \mathit{\#Ctor}_\sigma ~=~
  |\{ j \mid j \in \{1, \ldots, m\} \text{~and~} f_j : \sigma \}|
\end{equation*}
Similarly, each constructor~$f_i$ with $f_i : \sigma$ is given a unique
index~$\mathit{Id}_{f_i} \in \{1, \ldots, \mathit{\#Ctor}_\sigma\}$ as
identifier within its sort~$\sigma$:
\begin{equation*}
  \mathit{Id}_{f_i} ~=~ |\{ j \mid j \in \{1, \ldots, i\}
  \text{~and~} f_j : \sigma \}|
\end{equation*}
For each sort~$\sigma \in \{\sigma^d_1, \ldots, \sigma^d_k\}$, we
furthermore need to know the cardinality~$|\T_{\sigma}|$ of the
term domain~$\T_{\sigma}$. The cardinality can be derived by
computing the strongly connected components of the dependency graph
induced by the constructors (the graph with sorts $\sigma^d_1, \ldots,
\sigma^d_k$ as nodes, and edges $\sigma^d_i \to \sigma^d_j$ whenever there
is a constructor with a $\sigma^d_j$-sorted argument and result sort
$\sigma^d_i$). 
\Comment{Can we say it is infinite when graph has a loop?}
We write $|\T_{\sigma}| = \infty$ for infinite
domains.

\subsection{Definition of the Reduction}
\label{sec:reduction}

Suppose $\phi$ is a flat formula in negation normal form (NNF) over an
ADT as defined in Section~\ref{sec:prelims}. To translate $\phi$ to an
EUF+LIA formula~$\tilde \phi$, we introduce a new set of function
symbols ranging over integers: for each constructor~$f : \sigma_1
\times \cdots \times \sigma_n \to \sigma_0$ a new function~$\tilde f :
\Z^n \to \Z$ with the same arity~$n$; for each selector~$f^j :
\sigma_0 \to \sigma_j$ a unary function~$\tilde f^j : \Z \to \Z$; for
each sort~$\sigma \in \{\sigma^d_1, \ldots, \sigma^d_k\}$ a function
symbol~$\mathit{ctorId}_\sigma : \Z \to \Z$ to encode testers, and a
function~$\mathit{depth}_\sigma : \Z \to \Z$ to ensure acyclicity of
terms. Further, for each variable~$x : \sigma$ occurring in $\phi$, we
introduce an integer-valued variant~$\tilde x : \Z$.

The actual reduction is defined through the rewriting rules in the
upper half of Table~\ref{tab:reduction}. Since the reduction works
differently for positive and negative occurrences of
$\mathit{is}_f(x)$ literals, we assume that rules are only applied in
positive positions, and handle negation explicitly in the rules (and
assume that $\phi$ is in negation normal form). Rule~\eqref{rule:1}
augments every occurrence of a constructor symbol with corresponding
statements about selectors (ensuring that both are inverses of each
other); about the index $\mathit{Id}_f$ of the constructor (ensuring
that different constructors of the same sort produce distinct
values); and about the depth of the constructed term (ensuring that no
term can occur as sub-term of itself). Essentially the same
translation is done for testers by rule~\eqref{rule:3}, introducing
fresh constructor arguments through an existential quantifier.
Rule~\eqref{rule:2} augments each occurrence of a selector with a
disjunction stating that the considered term was actually created
using one of the constructors of the sort; this is necessary in
general since selectors~$f^j$ can be applied to terms constructed
using constructors other that $f$ (an optimisation is discussed in
Section~\ref{sec:opt}). Rule~\eqref{rule:4} asserts that the
constructor of a term is different from $f$, and \eqref{rule:5},
\eqref{rule:6} translate equations by simply renaming variables.

Suppose $\phi^*$ is the result of exhaustively applying the rules at
positive positions in $\phi$, and $x_1 : \sigma_1, \ldots, x_l :
\sigma_l$ are all variables occurring in $\phi$, then the reduct of
$\phi$ is defined as $ \tilde \phi = \phi^* \wedge \bigwedge_{i = 1}^l
\mathit{In}_{\sigma_i}(\tilde x_i)$.

\begin{example}
  \label{ex:encoded}
  In the encoded version of the formula from
  Example~\ref{ex:flattened}, all variables and functions range over
  integers; for readability, we keep the names of all variables. New
  variables~$\id{s1}, \ldots, \id{s4}$ are introduced to eliminate the
  quantifiers of $\mathit{ExCtorSpec}_f$ expressions through
  Skolemisation:
\begin{lstlisting}[escapechar=\%]
%\exLink{3}%  // encoding of x.is_cons
  cons(s1, s2) = x & ctorId_CList(x) = 1 &
  head(x) = s1 & tail(x) = s2 & 0<=s1 & s1<3 &
  depth_CList(x) > depth_Colour(s1) &
  depth_CList(x) > depth_CList(s2) &
  // encoding of blue = t1
  blue = t1 & ctorId_Colour(t1) = 2 &
  // encoding of y != t1 (unchanged)
  y != t1 &
    // encoding of red = t2
  ((red = t2 & ctorId_Colour(t2) = 0 &
    // encoding of x.head = t2
    head(x) = t2 & (
      // case x.is_nil
      (nil = x & ctorId_CList(x) = 0) |
      // case x.is_cons
      (cons(s3, s4) = x & ctorId_CList(x) = 1 &
       head(x) = s3 & tail(x) = s4 &
       0 <= s3 & s3 < 3 &
       depth_CList(x) > depth_Colour(s3) &
       depth_CList(x) > depth_CList(s4)))) |
    // encoding of nil = t3
   (nil = t3 & ctorId_CList(t3) = 0 &
    // encoding of cons(y, t3) = x
    cons(y, t3) = x & ctorId_CList(x) = 1 &
    head(x) = y & tail(x) = t3 &
    depth_CList(x) > depth_Colour(y) &
    depth_CList(x) > depth_CList(t3))) &
  // range constraints for x, y, t1, t2, t3
  // (some of which are just "true")
  0<=y & y<3 & 0<=t1 & t1<3 & 0<=t2 & t2<3
\end{lstlisting}
\end{example}

It should be noted that it is not necessary to assume positiveness of
the $\mathit{depth}_\sigma$ functions, since the functions are only
used to ensure acyclicity of terms by comparing the depth of a term
with the depths of its direct sub-terms. In general, although the
formula makes use of integer arithmetic, only very simple arithmetic
constraints are needed. Up to slight syntactic modifications, all
constraints fall into the Unit-Two-Variable-Per-Inequality fragment
UTVPI~\cite{DBLP:conf/ppcp/JaffarMSY94,DBLP:conf/cade/CimattiGS09},
i.e., only inequalities with up to two variables and unit coefficients
are needed. The constraints can therefore be both solved and
interpolated efficiently (of course, presence of Boolean structure or
negation still implies NP-hardness).


\iflong
\subsection{Correctness of Reduction}
\fi
%
\begin{theorem}
  \label{thm:reductionSoundComplete}
  The reduct~$\tilde \phi$ of a flat ADT formula~$\phi$ in NNF is
  satisfiable (over EUF+LIA) if and only if $\phi$ is satisfiable
  (over an ADT).
\end{theorem}

\iflong
\begin{IEEEproof}
  Since reduction preserves the Boolean structure of a formula, and
  the reduction rules are agnostic of the position at which they are
  applied, it is enough to prove the theorem for flat conjunctions of
  literals (i.e., formulas in negation normal form that do not contain
  disjunctions).

  ``$\Longleftarrow$'' (easy direction) Suppose $\phi$ is
  satisfiable, with structure~$(\T, I)$ and variable
  assignment~$\beta$. We construct a
  family~$(\alpha_{\sigma^d_i})_{i=1}^k$ of injective functions as
  embedding of the domains~$\T_{\sigma^d_i}$ into $\Z$. For $i$ such
  that $\T_{\sigma^d_i}$ is infinite, $\alpha_{\sigma^d_i}$ can be any
  bijection~$\T_{\sigma^d_i} \to \Z$; if $\T_{\sigma^d_i}$ is finite, we
  choose $\alpha_{\sigma^d_i}$ to be a
  bijection~$\T_{\sigma^d_i} \to \{0, \ldots, |\T_{\sigma^d_i}|-1\}$.  Let
  $\alpha = \bigcup_{i=1}^k \alpha_{\sigma^d_i}$. To satisfy
  $\tilde\phi$, choose variable
  assignment~$\tilde\beta = \alpha \circ \beta$, and the
  interpretation~$\tilde I$ of constructors and selectors over $\Z$
  that is induced by $\alpha$. Define
  $\tilde I(\mathit{depth}_\sigma)(n)$ to be the depth of the
  constructor term~$\alpha_\sigma^{-1}(n)$, and
  $\tilde I(\mathit{ctorId}_\sigma)(n)$ as the index~$\mathit{Id}_f$ of
  the head symbol~$f$ of $\alpha_\sigma^{-1}(n)$ (and arbitrary if
  $\alpha_\sigma^{-1}(n)$ is undefined).

  ``$\Longrightarrow$'' Suppose $\tilde\phi$ is satisfiable, with
  structure~$(\Z, \tilde I)$ and assignment~$\tilde \beta$. We
  construct a set~$P$ of relevant integer indices~$a$ and
  corresponding sorts~$\sigma$ in the model, and a mapping~$\gamma : P
  \to \T$ (with $\gamma(a, \sigma) \in \T_\sigma$ for each $(a,
  \sigma) \in P$) that can be used to define a variable
  assignment~$\beta(x : \sigma) = \gamma(\tilde\beta(\tilde x),
  \sigma)$ to satisfy $\phi$. The main difficulty is to ensure that
  $\gamma$ is injective, since otherwise disequalities in $\phi$ might
  be violated.

  We set $P = D \cup D_t$, where $D$ is the set of
  pairs~$(\tilde\beta(\tilde x), \sigma)$ for variables~$x : \sigma$
  for which $\phi$ contains a constructor literal~$f(\ldots) \approx
  x$, a selector literal~$f^j(x) \approx \ldots$, or a (possibly
  negated) tester~$\mathit{is}_f(x)$.  The encoding ensures that head
  symbols and children of terms represented by elements of $D$ are
  defined by the $\mathit{ctorId}_\sigma$ functions and the selectors;
  for $(a, \sigma) \in D$, define therefore $\mathit{dep}(a, \sigma) =
  \langle f, (c_1, \sigma_1), \ldots, (c_n, \sigma_n)\rangle$ if
  $\tilde I(\mathit{ctorId}_\sigma)(a) = \mathit{Id}_f$, with $f :
  \sigma_1 \times \cdots \times \sigma_n \to \sigma$, and $c_j =
  \tilde I(\tilde f^j)(a)$ for $j \in \{1, \ldots, n\}$.

  Let $D_t$ contain all pairs~$(c_i, \sigma_i)$ in tuples
  $\mathit{dep}(a, \sigma) = \langle f, (c_1, \sigma_1), \ldots, (c_n,
  \sigma_n)\rangle$, for any $(a, \sigma) \in D$; as well as
  pairs~$(\tilde\beta(\tilde x), \sigma) \not\in D$ for any further
  variable~$x : \sigma$ in $\phi$.

  We inductively define a sequence~$\gamma_0, \gamma_1, \ldots,
  \gamma_{|P|}$ of partial functions~$P \to \T$:
  \begin{enumerate}
  \item let $\gamma_0 = \emptyset$;
  \item for $i > 0$, if there is $(a, \sigma) \in D$ such that
    $\mathit{dep}(a, \sigma) = \langle f, (c_1, \sigma_1), \ldots, (c_n,
    \sigma_n)\rangle$,
    the function~$\gamma_{i-1}$ is defined for each
    pair~$(c_j, \sigma_j)$ (for $j \in \{1, \ldots, n\})$, but
    $\gamma_{i-1}$ is not defined for $(a, \sigma)$, then let
    $\gamma_i = \gamma_{i-1} \cup \{(a, \sigma) \mapsto
    f(\gamma_{i-1}(c_1, \sigma_1), \ldots, \gamma_{i-1}(c_n,
    \sigma_n))\}$.
  \item for $i > 0$, if case~2) does not apply, pick any pair~$(a,
    \sigma) \in P \setminus D$ for which $\gamma_{i-1}$ is not
    defined, and any constructor term~$s \in \T_\sigma$ that does not
    occur in the range of $\gamma_{i-1}$ yet; choose $(a, \sigma)$ and
    $s$ such that the \emph{depth of $s$ becomes minimal.}  Let
    $\gamma_i = \gamma_{i-1} \cup \{(a, \sigma) \mapsto s\}$.
  \end{enumerate}

  Importantly, the final function~$\gamma = \gamma_{|P|}$ is defined
  for all elements of $P$, and no two elements of $P$ are mapped to
  the same term. To see that $\gamma$ is defined for all elements of
  $P$, observe that the use of $\mathit{depth}_\sigma$ functions in
  the encoding ensures that the $\mathit{dep}$ function is acyclic,
  i.e., no term can ever be required to contain itself as a
  sub-term. To see that $\gamma$ is injective, observe that by
  definition the choice of $s$ in 3) cannot violate injectivity in
  $\gamma_i$. Different iterations of 2) cannot construct a term
  $f(\gamma_{i-1}(c_1, \sigma_1), \ldots, \gamma_{i-1}(c_n,
  \sigma_n))$ twice, due to the presence of constructor
  literals~$\tilde f(\ldots) \approx \tilde x$ in $\tilde\phi$ that
  are consistently interpreted.  Finally, the fact that case 2) is
  always preferred over 3) implies that the term $f(\gamma_{i-1}(c_1,
  \sigma_1), \ldots, \gamma_{i-1}(c_n, \sigma_n))$ has to contain the
  most recently chosen term~$s$ from case 3) (if there is any) as a
  sub-term; this implies that $f(\gamma_{i-1}(c_1, \sigma_1), \ldots,
  \gamma_{i-1}(c_n, \sigma_n))$ is deeper than all terms~$s$
  previously chosen in case~3), and therefore different from all of
  them.

  It is then possible to choose the variable assignment
  $\beta(x) = \gamma(\tilde\beta(\tilde x), \sigma)$ for each
  variable~$x : \sigma$ in $\phi$.
\end{IEEEproof}
\fi


\subsection{Two Optimisations}
\label{sec:opt}

The reduction, as presented so far, can be improved in a number of
ways. A first optimisation concerns the way selectors are
translated to EUF+LIA, rule~\eqref{rule:2}. It can be observed that
the disjunction of $\mathit{ExCtorSpec}_g$ literals introduced by
rule~\eqref{rule:2} is in most cases unnecessary, and usually the rule
can be simplified to
\begin{equation}
  \tag{\ref{rule:2}'}
  f^j(x) \approx y
  \quad\Longrightarrow\quad
  \tilde f^j(\tilde x) \approx \tilde y
\end{equation}
This simplification is possible whenever rule~\eqref{rule:2} is
applied to \emph{guarded} selector literals, i.e., whenever
$f^j(x) \approx y$ occurs in conjunction with a (positive or negative)
test~$\mathit{is}_g(x)$, or in conjunction with a constructor
literal~$g(x_1, \ldots, x_n) \approx x$ (in both cases, regardless of
whether $f = g$).

\begin{example}
  The effect of this redundancy can be seen in
  Example~\ref{ex:encoded}: given lines~1--5, the disjunction in
  14--21 can be simplified to \mbox{\lstinline!s3
    = s1 & s4 = s2!}, and can be removed entirely since \id{s3} and
  \id{s4} do not occur elsewhere in the formula.
  \hfill\IEEEQED
\end{example}

\begin{example}
  The full rule~\eqref{rule:2} is necessary for the following formula
  over the ADT in Example~\ref{ex:lists}:
  \begin{lstlisting}[escapechar=\%]
%\exLink{5}%  x = cons(x.head, x.tail) <-> x.is_cons
  \end{lstlisting}
  This is because \lstinline!x.head! and \lstinline!x.tail! occur
  unguarded.
  \hfill\IEEEQED
\end{example}

As a second optimisation, the treatment of sorts with finite domain
can be improved, in particular for sorts that are \emph{enumerations}
(i.e., sorts with only nullary constructors). The full EUF encoding is
overkill for enumerations, since instead we can map each
constructor~$f$ directly to the index~$\mathit{Id}_{f}$:
\begin{equation}
  \tag{\ref{rule:1}'}
  f \approx x_0
  \quad\Longrightarrow\quad
  \mathit{Id}_f \approx \tilde x_0
\end{equation}
Similarly, testers in enumerations reduce to simple arithmetic
comparisons.


\subsection{Size Increase Caused by the Reduction}
The reduction rules replace every literal in a formula~$\phi$ with an
expression that is linear in the size~$n$ of the considered ADT, so
that $|\tilde \phi| \in O(n \cdot |\phi|)$. If the ADT is considered
as fixed, the reduction is linear.

As an experimental evaluation of the size increase, we applied the
procedure (including the optimisations from the previous section) to
the 8000 randomly generated ADT benchmarks from
\cite{DBLP:journals/jsat/BarrettST07} (4422 of the benchmarks are
unsat). The benchmarks themselves are not very challenging, with the
most complicated one solved in around~1\,s, and the average solving
time of 43\,ms dominated by parsing, pre-processing, etc. The average
problem sizes, counted as the number of sub-expressions of each
formula, were:
\begin{center}
  \begin{tabular}{ccc}
    After parsing & After reduction & After red.
    \& simpl.
    \\\hline
    76 & 337 & 34
  \end{tabular}
\end{center}
This means that reduction led to an increase in size by a factor of
4.5, but this increase was more than offset by subsequent
simplification (using the standard EUF+LIA simplifier implemented in
\textsc{Princess}). Analysing further, it turned out that reduction
followed by simplification was extremely effective on the
unsatisfiable benchmarks: of the 4422 unsatisfiable problems, 4334
could directly be simplified to $\mathit{false}$. The average size of
the remaining 3666 problems, after reduction and simplification, was
74, incidentally the same as the average initial size of all
benchmarks.

An experimental comparison of our solver with other SMT solvers, on a
larger set of benchmarks, is ongoing.

\section{Craig Interpolation in (Extensions of) ADTs}
\label{sec:interpolation}

Since quantifier-free Craig interpolation in EUF+LIA is well
understood (e.g.,
\cite{vmcaiInterpolation2011,DBLP:conf/spin/ChristHN12,DBLP:conf/cade/CimattiGS09}),
the reduction approach can also be leveraged for interpolation. Given
an unsatisfiable conjunction~$\phi_A \wedge \phi_B$, the problem of
(reverse) interpolation is to find a formula~$I$ such that~$\phi_A
\Rightarrow I$, $\phi_B \Rightarrow \neg I$, and all variables in $I$
are common to $\phi_A$ and $\phi_B$. If $\phi_A, \phi_B$ are ADT
formulas, it is natural to approach interpolation by first computing
an EUF+LIA interpolant~$\tilde I$ for the reduced conjunction $\tilde
\phi_A \wedge \tilde \phi_B$.

\begin{example}
  An interpolation problem over the list ADT from
  Example~\ref{ex:lists} is:
\begin{lstlisting}[escapechar=\%]
%\exLink{6}%  \part[left]  (x.is_cons & x.tail = z &
                z.is_cons & x.head != z.head)
& \part[right] (x = cons(c, cons(c, y)))
\end{lstlisting}
The only common variable of the two formulas is \id{x}, and a solution
of the interpolation problem is the disequality
\lstinline?x.head != x.tail.head?.
Note that this formula is a correct interpolant even though the
selectors are unguarded.
\hfill\IEEEQED
\end{example}

To translate~$\tilde I$ back to an ADT interpolant~$I$, three main
points have to be addressed. First, since all ADT sorts are translated
to integers, the formula~$\tilde I$ might contain arithmetic
operations on ADT terms that cannot easily be mapped back to the ADT
world. This turns out to be a non-issue for infinite ADT sorts, since
reduction does not make use of arithmetic operations for terms over
infinite sorts (indeed, equivalently every infinite ADT
sort~$\sigma^d$ could be mapped to a fresh uninterpreted
sort~$\tilde\sigma^d$). The situation is different for finite sorts,
where predicates~$\mathit{In}_\sigma$ from
Table~\ref{tab:reduction} represent cardinality constraints that can
contribute to unsatisfiability of a formula. One solution is the
optimisation discussed in Section~\ref{sec:opt}: by defining a
\emph{fixed} mapping of terms in finite domains to integers,
translation of interpolants back to ADT formulas is significantly
simplified.\iflong\footnote{In our implementation, such fixed mapping is
  currently only done for enumerations, not for other finite ADT
  sorts.}\fi

Second, the functions~$\mathit{ctorId}_\sigma$ introduced by the
reduction are not valid ADT operations, and have to be translated back
to testers (which can be done quite easily).

Third, interpolants might also mention $\mathit{depth}_\sigma$
operations, which have no direct correspondence in the original ADTs
theory. Instead of devising ways how to eliminate such operations, we
decide to embrace them instead as a useful extension of ADTs, and
adapt our reduction method accordingly. Since depth is but one measure
that can be used to ensure acyclicity, the next sections therefore
discuss how we can reason about ADTs with \emph{size constraints.}


\section{Solving ADTs with Size Constraints\label{sect::size}}

We now consider ADT formulas extended with constraints about term
size. The \emph{size}~$|t|$ of a term~$t \in \T$ is the number of
constructor occurrences in $t$. The resulting formal language is an
extension of the language defined in Section~\ref{sec:prelims}:
\begin{align*}
  \phi ~~::=~ & \ldots
  ~|~~ \phi_{\text{Pres}}(|t_1|, \ldots, |t_n|)
                           &&
                              \text{Size constraints}
\end{align*}
where $\phi_{\text{Pres}}(|t_1|, \ldots, |t_n|)$ is any Presburger
formula about the size of ADT terms~$t_1, \ldots, t_n$.

\begin{example}
  Consider the ADT in Example~\ref{ex:lists}, and a variable $x$ of
  sort~\id{CList}. The formula
  \begin{lstlisting}[escapechar=\%]
%\exLink{7a}%  \size(x) = 3 & x.head = blue
  \end{lstlisting}
  has the satisfying assignment
  $x \mapsto \id{cons}(\id{blue}, \id{nil})$, and this assignment is
  unique. In contrast, the formula
  \begin{lstlisting}[escapechar=\&]
&\exLink{7b}&  \size(x) % 2 = 0
  \end{lstlisting}
  is unsatisfiable, since the size of any list term is odd (term size
  does not exactly coincide with the length of a list).
  \hfill\IEEEQED
\end{example}

To extend our reduction approach to formulas with size constraints,
there are two main issues that have to be addressed: (i) constructor
terms might not exist for all sizes~$n \in \N_{\geq 1}$, and (ii) even
if terms of some size~$n \in \N_{\geq 1}$ exist, there might be too
few of them to satisfy a formula.

\begin{example}
  \label{ex:nats}
  Consider the ADT of positive natural numbers:
  \begin{lstlisting}
\sorts {
  Nat { one; succ(Nat pred); };
}
  \end{lstlisting}
  For every size~$b \in \N_{\geq 1}$ there is exactly one constructor
  term~$t$ with $|t| = b$. This implies unsat.\ of the formula
  \begin{lstlisting}[escapechar=\%]
%\exLink{8}%  \size(x) = 3 & \size(y) = 3 & x != y
  \end{lstlisting}
%
  %
\end{example}


\subsection{Reduction and Incremental Unfolding}

\begin{table*}[t]
  \fboxsep2mm
  \fbox{
    \begin{minipage}{\linewidth-2\fboxsep}
      Additional reduction rule for size expressions, assuming
      $x$ is a variable of sort~$\sigma \in \{\sigma^d_1, \ldots, \sigma^d_k\}$,
      and $y$ a variable of sort~$\Z$:
      \begin{align}
        \label{rule:7}
        |x| \approx y
        \quad\Longrightarrow\quad &
        \mathit{size}_{\sigma}(\tilde x) \approx y ~\wedge~
        y \in \BBS_{\sigma}
      \end{align}

      Compared to Table~\ref{tab:reduction}, for each
      constructor $f : \sigma_1 \times \cdots \times \sigma_n \to \sigma_0$
      the abbreviation $\mathit{CtorSpec}_f$ is replaced with
      $\mathit{CtorSpec}'_f$:
      \begin{align*}
        \mathit{CtorSpec}'_f(x_0, \ldots, x_n)
        \quad=\quad &
     \left(
      \begin{array}{@{}>{\displaystyle}l@{}}
        \tilde f(x_1, \ldots, x_n) \approx x_0 ~\wedge~
        \mathit{ctorId}_{\sigma_0}(x_0) \approx \mathit{Id}_f ~\wedge
        \\
        \bigwedge_{j = 1}^n \big(
        \tilde f^j(x_0) \approx x_j ~\wedge~
        \mathit{size}_{\sigma_j}(x_j) \in \BBS_{\sigma_j}
        \big) ~\wedge~
        \mathit{size}_{\sigma_0}(x_0) \approx
        1 + \sum_{j=1}^n \mathit{size}_{\sigma_j}(x_j)
    \end{array}
    \right)
  \end{align*}
    \end{minipage}}

  \caption{Additional rules for reduction of ADTs with size
    constraints to EUF+LIA}
  \label{tab:reductionSize}
\end{table*}

We address issue~(i) noted above by reasoning globally about possible
term sizes within an ADT. For an ADT sort~$\sigma^d_j$, we define
$\BBS_{\sigma^d_j} = \{ |t| \mid t \in \T_{\sigma^d_j} \} \subseteq
\N$ to be the \emph{size image} of the term set~$\T_{\sigma^d_j}$,
i.e., the set of term sizes in $\T_{\sigma^d_j}$. The size image turns
out to be a special case of the \emph{Parikh image} of a context-free
language, since an ADT can be interpreted as a context-free grammar
over a singleton alphabet (by considering every sort as a non-terminal
symbol, and mapping every constructor to the unique letter in the
singleton alphabet). This implies that $\BBS_{\sigma^d_j}$ is
semi-linear, and that a representation of the set in the form of an
existential Presburger formula can be derived from the set of
constructors in linear time~\cite{DBLP:conf/cade/VermaSS05}.

\Comment{no longer UTVPI; discuss implications for interpolation}
Table~\ref{tab:reductionSize} shows how the reduction from
Section~\ref{sec:reduction} (and Table~\ref{tab:reduction}) is
augmented to deal with size constraints. Instead of the
$\mathit{depth}_\sigma$ functions, for each
sort~$\sigma \in \{\sigma^d_1, \ldots, \sigma^d_k\}$ a
function~$\mathit{size}_\sigma : \Z \to \Z$ representing term size is
introduced, and the $\mathit{CtorSpec}_f$ constraints are changed
accordingly; and an additional reduction rule~\eqref{rule:7} is
introduced to handle equations~$|x| \approx y$ with the size
operation. Rule~\eqref{rule:7} adds constraints~$y \in \BBS_{\sigma}$
to ensure that only genuine term sizes are considered, assuming
implicitly that the size image~$\BBS_{\sigma}$ is represented as a
Presburger formula.

The resulting modified reduction approach is sound for checking
unsatisfiability of ADT formulas:
\begin{lemma}
  \label{lem:sizeUnsat}
  If the reduct~$\tilde \phi$ of a flat ADT formula~$\phi$ in NNF with
  size constraints is unsatisfiable, then $\phi$ is unsatisfiable.
\end{lemma}


%
Reduction does not directly give rise to a decision procedure for ADT
constraints with size constraints, in contrast to the situation
without size. This is because reduction does not precisely translate the
\emph{number} of terms for each size~$n \in \N_{\geq 1}$
(issue~(ii) from above). We can observe that the reduct~$\tilde\phi$
of the formula~$\phi$ in Example~\ref{ex:nats} is satisfiable, while
$\phi$ is unsatisfiable, showing that reduction alone is \emph{not}
sound for satisfiability (unsurprisingly).

Different approaches exist to establish soundness also for
satisfiability, in particular the extraction of \emph{length
  constraint completion} formulas~\cite{DBLP:journals/iandc/ZhangSM06}
that precisely define term sizes with sufficiently many distinct
terms. We follow the approach of incrementally unfolding (aka.\
unrolling) from \cite{Suter:2010:DPA:1707801.1706325}, which is quite
flexible, and complete in many relevant cases.

Let $\phi$ again be a (flat and NNF) ADT formula with size constraints. We
construct unfolding sequences~$\phi_0, \phi_1, \ldots$ by setting
$\phi_0 = \phi$, and for each $i \geq 1$ deriving $\phi_i$ by
unfolding one ADT variable~$x : \sigma$ that occurs in
$\phi_{i-1}$. If $f_1, \ldots, f_n$ are all constructors of the
considered ADT, we set
\begin{equation*}
  \phi_i ~~=~~ \phi_{i-1} \wedge
  \bigvee_{\substack{j \in \{1, \ldots, n\} \\ f_j : \sigma}}
   f_j(x_1^j, x_2^j, \ldots) \approx x
\end{equation*}
with fresh sorted argument variables~$x_1^1, x_2^1, \ldots, x_1^2,
x_2^2, \ldots$.

In practice, unfolding will usually happen incrementally: the next
variable to unfold is selected based on a model of the previous
partial unfolding~$\phi_{i-1}$, until enough terms have been
constructed to obtain a genuine model of $\phi$, or unsatisfiability
is detected.
\begin{lemma}
  \label{lem:unfolding}
  Let $\phi_0, \phi_1, \ldots$ be an unfolding sequence for $\phi$,
  and for each $i \in \N$ let $U_i$ be the set of variables unfolded
  in $\phi_i$ (i.e., $U_0 = \emptyset$, and $U_i = U_{i-1} \cup \{x\}$
  if $\phi_i$ was derived by unfolding $x$ in $\phi_{i-1}$). Then for
  any $i \in \N$:
  \begin{enumerate}
  \item if $\tilde \phi_i$ is unsatisfiable (over EUF+LIA) then $\phi$
    is unsatisfiable (over ADTs with size);
  \item if $\tilde \phi_i$ is satisfied by a model~$\tilde M$ and
    assignment~$\tilde \beta$, such that for every ADT variable~$x :
    \sigma$ in $\phi_i$ there is a variable~$y \in U_i$ with $y :
    \sigma$ and $\val_{\tilde M, \tilde \beta}(\tilde x) =
    \val_{\tilde M, \tilde \beta}(\tilde y)$, then $\phi$ is
    satisfiable (over ADTs with size).
  \end{enumerate}
\end{lemma}

\iflong
\begin{IEEEproof}
  1) follows directly from Lemma~\ref{lem:sizeUnsat}.
  
  2) Models over EUF+LIA can be translated to ADT models like in the
  proof of Theorem~\ref{thm:reductionSoundComplete}
  ``$\Longrightarrow$''. It can be noted that case~3) in the proof
  never applies due to the assumption that all variables are mapped to
  unfolded terms.
\end{IEEEproof}
\fi

\begin{example}
  In Example~\ref{ex:nats}, unsatisfiability is detected after unfolding
  $x$ and $y$ three times each.
  \hfill\IEEEQED
\end{example}

As the next example shows, however, unfolding is not always enough to
show unsatisfiability of a formula. The next sections will therefore
formulate a sufficient and necessary criterion for termination of
unfolding.
\begin{example}
  \label{ex:nats2}
  With the ADT from Example~\ref{ex:nats}, the formula
  \begin{lstlisting}[escapechar=\%]
%\exLink{10}%  \size(x) = \size(y) & x != y
  \end{lstlisting}
  is unsatisfiable, but cannot be shown to be unsatisfiable with a
  finite number of unfolding steps.
  \hfill\IEEEQED
\end{example}


\subsection{Completeness and Incompleteness of Unfolding}

We give a precise characterisation of the ADTs for which unfolding
will allow us to eventually detect (un)satisfiable of a formula, and
therefore gives rise to a decision procedure.  As identified
in\cite{Suter:2010:DPA:1707801.1706325}, the essential property of an
ADT (resp., of the considered catamorphism, which in our case is the
size function) is \emph{sufficient surjectivity}, implying that ADTs
are sufficiently populated to satisfy disequalities in a formula: the
number of terms of size~$b$ grows unboundedly when $b$ tends to
infinity. We write $\T^k_{\sigma}$ for the set of constructor terms of
ADT sort~$\sigma$ and size~$k$, i.e., $\T^k_{\sigma} = \{ t \in
\T^k_{\sigma} \mid |t| = k\}$.

\begin{definition}
  An ADT sort~$\sigma^d$ is \emph{expanding} if for every natural
  number~$n \in \N$ there is a bound~$b \in \N$ such that for
  every~$b' \geq b$ either $\T^{b'}_{\sigma^d} = \emptyset$ or
  $|\T^{b'}_{\sigma^d}| \geq n$. An ADT is expanding if each of its
  sorts~$\sigma^d_1, \ldots, \sigma^d_k$ is expanding.
\end{definition}

\begin{example}
  An example of an ADT that is \emph{not} expanding are the
  natural numbers (Example~\ref{ex:nats}): for every size~$b \in
  \N_{\geq 1}$ there is exactly one constructor term~$t$ with $|t| =
  b$.
  \hfill\IEEEQED
\end{example}

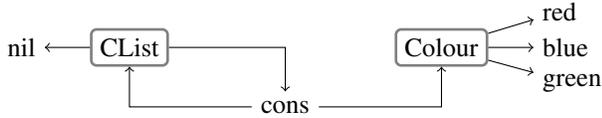
\begin{figure}[t]
  \centering
  \begin{tikzpicture}
    \path[use as bounding box] (-0.5,-0.8) rectangle(5,0.7);

    \node[depSort] (clist) {CList};
    \node[depSort,right=3 of clist] (col) {Colour};
    \node[depCtor,left=0.6 of clist] (nil) {nil};
    \node[depCtor,right=0.6 of col,yshift=-3ex] (green) {green};
    \node[depCtor,right=0.6 of col,yshift=3ex] (red) {red};
    \node[depCtor,right=0.6 of col] (blue) {blue};
    \node[depCtor,right=1.1 of clist,yshift=-5ex] (cons) {cons};

    \draw[->] (clist) -- (nil);
    \draw[->] (col) -- (red);
    \draw[->] (col) -- (green);
    \draw[->] (col) -- (blue);
    \draw[->] (clist) -| (cons);
    \draw[->] (cons) -| (clist);
    \draw[->] (cons) -| (col);
  \end{tikzpicture}

  \caption{Dependency graph for the list ADT from Example~\ref{ex:lists}}
  \label{fig:depGraph}
\end{figure}

\begin{lemma}
  Systematic unfolding terminates (i.e., in every unfolding
  sequence~$\phi_0, \phi_1, \ldots$ in which every variable is
  eventually unfolded, eventually one of the cases of
  Lemma~\ref{lem:unfolding} applies) for all formulas~$\phi$ if and
  only if the considered ADT is expanding.
\end{lemma}

\iflong
\begin{IEEEproof}[Proof sketch]
  ``$\Longrightarrow$'' 
  Example~\ref{ex:nats2} generalises to arbitrary non-expanding ADTs:
  for every non-expanding sort~$\sigma$ there is a constant~$c \in \N$
  and an infinite semi-linear set~$S \subseteq \BBS_\sigma$ such that
  $|\T^{b}_{\sigma}| < c$ for all $b \in S$. The existence of $c, S$
  follows from the proof of Theorem~\ref{thm:expanding} below.

  ``$\Longleftarrow$'' 
  Consider first the case of $\phi$ being a conjunction of
  disequalities and a size constraint~$\phi_{\text{Pres}}(|x_1|,
  \ldots, |x_n|)$. Since the ADT is expanding, satisfiability of
  $\phi$ reduces to the question whether the size images of the
  $x_i$-domains contain elements large enough, and compatible with
  $\phi_{\text{Pres}}$, that all disequalities can be
  satisfied. Systematic unfolding of $x_1, \ldots, x_n$ will add size
  image constraints~$|x'| \in \BBS_\sigma$ for all sub-terms, and
  either find a set of satisfying term sizes (and corresponding
  terms), or conclude unsatisfiability because the conjunction of size
  images and size constraint~$\phi_{\text{Pres}}$ becomes
  inconsistent.

  Adding constructor, selector, or test literals does not change the
  argument, since solutions of such literals can be represented in the
  form of a most-general unifier~\cite{Suter:2010:DPA:1707801.1706325}.
\end{IEEEproof}

\medskip
\fi
Non-expandingness turns out to be a corner case: all non-expanding
ADTs more or less look like the natural numbers
(Example~\ref{ex:nats}), and most other practical ADTs are expanding.
For instance, both ADT sorts in Example~\ref{ex:lists} expand.


\subsection{Effective Characterisation of Expanding ADTs}

To characterise expanding ADTs, we first make the simplifying
assumption that all ADT sorts~$\sigma^d_j$ contain at least two
constructor terms; sorts with only a single term can obviously be
eliminated easily from a constraint.  As a further piece of notation,
we need a relativised version of the size image: for an ADT
sort~$\sigma^d_j$ and a constructor~$f$, we write
\begin{equation*}
  \BBS_{\sigma^d_j}^f = \{ |t| \mid t \in \T_{\sigma^d_j}, \text{and~}
  t \text{~does not start with~} f  \}
\end{equation*}
for the size image restricted to terms with head symbol $\not= f$.

Consider then the bipartite dependency graph~$D = (V, E)$ with
vertices~$V = \{\sigma^d_1, \ldots, \sigma^d_k\} \cup \{f_1, \ldots,
f_m\}$ being sorts and constructors, and the edge set
\begin{equation*}
  E ~=~ \left\{
  \begin{array}{@{}l@{}}
    (\sigma_0, f_j), (f_j, \sigma_1), \ldots, (f_j, \sigma_n)\\
    \qquad
    \mid j \in \{1, \ldots, m\},
    f_j: \sigma_1 \times \cdots \times \sigma_n \to \sigma_0
  \end{array}
  \right\}
\end{equation*}
Fig.~\ref{fig:depGraph} gives the graph for the list ADT in
Example~\ref{ex:lists}.

\begin{theorem}
  \label{thm:expanding}
  An ADT is \emph{not} expanding if and only if the graph~$D$ contains
  a simple
  cycle~$C = \sigma^1 \to f^1 \to \sigma^2 \to f^2 \to \cdots \to f^n
  \to \sigma^1$ with the following properties:
  \begin{enumerate}
  \item $C$ is the only path from $\sigma^1$ to itself, i.e., every
    cycle starting and ending in $\sigma^1$ is a repetition of $C$;
  \item all constructors $f^1, f^2, \ldots, f^n$ on $C$ are unary;
  \item the cycle~$C$ unboundedly contributes to the size
    image~$\BBS_{\sigma^1}$, i.e.,
    \begin{equation*}
      \forall k.~~ \BBS_{\sigma^1} \not=
      \{0, \ldots, k\} \cdot n +
      \bigcup_{i = 1}^{n} \big( \BBS_{\sigma^{i}}^{f^{i}} + i - 1  \big)~.
    \end{equation*}
  \end{enumerate}
\end{theorem}

The characterisation theorem implies that every non-expanding ADT has
a set of cyclically connected sorts~\id{S1}, \ldots, \id{Sn}, each of
which might contain further constructors \lstinline!c1_1!,
\lstinline!c1_2!, \ldots\ that do not lead back to \id{S1}, \ldots,
\id{Sn}:
\begin{lstlisting}
\sorts {
  // ...
  S1 { f1(S2 s2); c1_1; c1_2; /* ... */ };
  S2 { f2(S3 s3); c2_1; c2_2; /* ... */ };
  // ...
  Sn { fn(S1 s1); cn_1; cn_2; /* ... */ };
  // ...
}
\end{lstlisting}

The conditions of the theorem are clearly satisfied for the ADT of
natural numbers (Example~\ref{ex:nats}). Condition~3)
is satisfied whenever $\BBS_{\sigma^{i}}^{f^{i}}$ is finite for all
$i \in \{1, \ldots, n\}$, but there are more subtle situations\iflong:

\begin{example}
  We extend the list ADT from Example~\ref{ex:lists} by adding two
  further sorts:
  \begin{lstlisting}
\sorts {
  S1 { f1(S2 s2); };
  S2 { f2(S1 s1); null; col(CList list); };
}
  \end{lstlisting}
  The domain of sort~\id{S1} contains terms of any size greater than
  one. However, while the number of terms of size~$2k + 1$ grows
  exponentially with $k$, there is exactly one term of size~$2k$ for
  every $k$, proving non-expandingness.

  A cycle of length~3, in contrast, yields an expanding ADT:
  \begin{lstlisting}
\sorts {
  S1 { f1(S2 s2); };
  S2 { f2(S3 s3); };
  S3 { f3(S1 s1); null; col(CList list); };
}
  \end{lstlisting}
  We can note that condition~3) of Theorem~\ref{thm:expanding} now
  fails.
  \hfill\IEEEQED
\end{example}

Before we can prove Theorem~\ref{thm:expanding}, we need some further
results about ADTs. Consider constructor terms~$t[\bullet]$ with a
unique hole~$\bullet$; such terms can alternatively be seen as terms
with a single occurrence of a sorted variable. Composition of terms
with holes is defined as $t_1[\bullet] \circ t_2[\bullet] =
t_1[t_2[\bullet]]$. Two terms~$t_1[\bullet], t_2[\bullet]$ with holes
are \emph{incomparable} if $t_1[s_1] \not= t_2[s_2]$ for all
constructor terms~$s_1, s_2$ of the right sort. The
size~$|t[\bullet]|$ of a term with hole is the number of constructor
symbol occurrences in $t$, i.e., the hole does not count. This implies
$|t_1[\bullet] \circ t_2[\bullet]| = |t_1[\bullet]| + |t_2[\bullet]|$.
\begin{lemma}
  \label{lem:incompTerms}
  Suppose an ADT sort~$\sigma$ contains two incomparable constructor
  terms~$t_1[\bullet], t_2[\bullet]$ with holes~$\bullet$ of
  sort~$\sigma$. Then $\sigma$ expands.
\end{lemma}

\begin{IEEEproof}
  Fix some $n \in N$. We need to show that there is a bound~$b \in \N$
  ensuring $\T^{b'}_{\sigma} = \emptyset$ or $|\T^{b'}_{\sigma}| \geq
  n$ for every $b' \geq b$. Observe that for every $g \geq n$ there
  are $\mbox{}>n$ pairwise incomparable terms with holes $t_1^0 \circ
  t_2^g \circ t_1^{g},~ t_1^1 \circ t_2^g \circ t_1^{g-1}, \ldots,
  t_1^g \circ t_2^g \circ t_1^{0}$, all of which have size~$g \cdot
  (|t_1| + |t_2|)$.  The size image~$\BBS_\sigma \subseteq \N$ can be
  represented as a finite union of arithmetic progressions,
  $\BBS_\sigma = \bigcup_{i=1}^l \{ a_i + k \cdot b_i \mid k \in \N
  \}$ with $a_i, b_i \in \N$ for $i \in \{1, \ldots, l\}$. For each $i
  \in \{1, \ldots, l\}$ with $b_i > 0$, assuming that $k \geq n \cdot
  (|t_1| + |t_2|)$ we can then pick $g = b_i \cdot n$, and some term
  $t : \sigma$ of size $a_i + (k - n \cdot (|t_1| + |t_2|)) \cdot
  b_i$, and obtain $\mbox{}>n$ pairwise distinct terms
  \begin{equation*}
    (t_1^0 \circ t_2^g \circ t_1^{g})[t],~
    (t_1^1 \circ t_2^g \circ t_1^{g-1})[t],~
    \ldots,~
    (t_1^g \circ t_2^g \circ t_1^0)[t]
  \end{equation*}
  that are all of size $a_i + (k - n \cdot (|t_1| + |t_2|)) \cdot b_i
  + g \cdot (|t_1| + |t_2|) = a_i + k \cdot b_i$.
  This implies that there is also a global bound~$b \in \N$ such that
  $\T^{b'}_{\sigma} = \emptyset$ or $|\T^{b'}_{\sigma}| \geq n$ for
  $b' \geq b$.
\end{IEEEproof}

\medskip
\begin{IEEEproof}[Proof of Theorem~\ref{thm:expanding}]
  ``$\Longrightarrow$'' 
  Suppose an ADT is not expanding, which means that there is a
  non-expanding sort~$\sigma$. Choose $\sigma$ such that whenever
  $\sigma \stackrel{*}{\to} \sigma'$ in the $D$-graph, and $\sigma'$
  is non-expanding as well, then $\sigma'$ is in the same strongly
  connected component (SCC) as $\sigma$; this is possible because the
  SCCs of $D$ form a DAG. Then $\BBS_\sigma$ has to be infinite
  (otherwise $\sigma$ would be expanding), and there is a simple
  $D$-path $C = \sigma^1 \to f^1 \to \sigma^2 \to f^2 \to \cdots \to
  f^n \to \sigma^1$ with $\sigma^1 = \sigma$ (otherwise there would be
  a non-expanding sort~$\sigma'$ reachable from $\sigma$, but not in
  the same SCC).

  We show that $C$ satisfies the conditions of the theorem. 1) holds
  because if there was a second path $C'$ from $\sigma$ to itself that
  is not a repetition of $C$, then both paths could be translated to
  incomparable $\sigma$-terms~$t_1[\bullet], t_2[\bullet]$ with
  $\sigma$-sorted holes~$\bullet$, and by Lemma~\ref{lem:incompTerms}
  the sort~$\sigma$ would be expanding. The same argument implies that
  2) holds: if any of the constructors $f_i$ had multiple arguments,
  incomparable terms $t_1[\bullet], t_2[\bullet]$ could be derived
  (since, by assumption, every sort contains at least two constructor
  terms).

  Suppose finally that 3) does not hold, i.e., for some $k \in \N$
  \begin{equation*}
    \BBS_{\sigma} ~=~
    \{0, \ldots, k\} \cdot n +
    \bigcup_{i = 1}^{n} \big( \BBS_{\sigma^{i}}^{f^{i}} + i - 1  \big)
  \end{equation*}
  This would imply that from some sort~$\sigma^{i}$ a non-expanding
  sort~$\sigma'$ is reachable, by following a constructor other
  than~$f^i$. By choice of $\sigma$, then $\sigma'$ has to be in the
  same SCC as $\sigma$, therefore there is a path from $\sigma'$ to
  $\sigma$, and condition 1) would be violated.

  ``$\Longleftarrow$'' 
  Suppose there is a cycle~$C$ satisfying 1)--3). Note that due to 1)
  and 2) we have the equality
  \begin{equation*}
    \BBS_{\sigma} ~=~
    \N \cdot n +
    \underbrace{\bigcup_{i = 1}^{n} \big( \BBS_{\sigma^{i}}^{f^{i}} + i - 1
      \big)}_R
  \end{equation*}
  Together with 3), this means $\N \cdot n + R \not= \{0, \ldots, k\}
  \cdot n + R$ for every $k \in \N$. Because $R$ is a semi-linear set,
  then there has to be a finite subset~$S \subseteq R$ such that for
  infinitely many points~$s \in \BBS_{\sigma}$ we have $\{ x \in R
  \mid s \in \N \cdot n + x \} \subseteq S$. Because the set $\{ t \in
  \T_\sigma \mid |t| \in S\}$ of terms is finite,\footnote{This
    property breaks down when ADTs are combined with other infinite
    data types, e.g., lists over $\Z$. In this case condition~3) has
    to be modified.} this immediately implies that the sort~$\sigma$
  is not expanding.
\end{IEEEproof}
\else
{} with infinite domains.
\fi





\section{Conclusions}

At the moment we are exploring applications and further extensions of
our approach. We are in the process of integrating our procedure into
the model checker \textsc{Eldarica}~\cite{Rummer13} to handle implication
checks and interpolation for ADTs; this also requires combination with
other data types, and in the long run likely interpolation
heuristics. It is also frequently necessary to combine ADTs with
quantifier reasoning and recursively defined functions, a direction
that requires further work. Finally, as a side-effect of
Theorem~\ref{thm:expanding}, there is a simple way to achieve
termination also for non-expanding ADTs, namely by replacing the cycle
with an explicit counter ranging over a built-in type of natural
numbers.

\paragraph{Acknowledgements}

    R\"ummer was supported by the
    Swedish Research Council
    under grant 2014-5484.

\bibliographystyle{plain}
\bibliography{refs}

\end{document}